\documentclass[letterpaper,preprintnumbers,prd,twocolumn,nofootinbib,nobibnotes,showpacs]{revtex4}
\usepackage{graphics,epsfig,subfigure}
\usepackage{color}
\usepackage{amsfonts}
\usepackage{mathrsfs}
\usepackage{epsfig}
\usepackage{graphicx}%
\usepackage{dcolumn}
\usepackage{amsmath}
\usepackage{color}
\usepackage{color}
\usepackage{graphics,epsfig,subfigure}
\usepackage{color}
\usepackage{amsfonts}
\usepackage{mathrsfs}
\usepackage{epsfig}
\usepackage{graphicx}%
\usepackage{dcolumn}
\usepackage{amsmath}
\usepackage{color}
\usepackage{color}
\begin{document}
\newcommand{\red}[1]{\textcolor{red}{{#1}}}
\newcommand{\blue}[1]{\textcolor{blue}{{#1}}}

\renewcommand{\baselinestretch}{1.3}
\newcommand\beq{\begin{equation}}
\newcommand\eeq{\end{equation}}
\newcommand\beqn{\begin{eqnarray}}
\newcommand\eeqn{\end{eqnarray}}
\newcommand\nn{\nonumber}
\newcommand\fc{\frac}
\newcommand\lt{\left}
\newcommand\rt{\right}
\newcommand\pt{\partial}

\allowdisplaybreaks

\title{A novel two-field pure K-essence for inflation, dark matter, dark energy and black holes}
\author{Changjun Gao\footnote{gaocj@nao.cas.cn}}

\affiliation{National Astronomical Observatories, Chinese Academy of Sciences, 20A Datun Road, Beijing 100101, China}

\affiliation{School of Astronomy and Space Sciences, University of Chinese Academy of Sciences,
19A Yuquan Road, Beijing 100049, China}

\begin{abstract}
K-essence theories are usually studied in the framework of one scalar field $\phi$. Namely, the Lagrangian of K-essence is the function of scalar field $\phi$  and its covariant derivative. However, in this paper, we explore a two-field pure K-essence, i.e. the corresponding Lagrangian is the function of covariant derivatives of two scalar fields without the dependency of scalar fields themselves. That is why we call it pure K-essence. The novelty of this K-essence is that its Lagrangian contains the quotient term of the kinetic energies from the two scalar fields. This results in the presence of many interesting features, for example, the equation of state can be arbitrarily small and arbitrarily  large. As a comparison, the range for equation of state of quintessence is from $-1$ to $+1$. Interestingly, this novel K-essence can play the role of inflation field, dark matter and dark energy. Finally, the absence of the scalar fields themselves in the equations of motion makes the study considerable simple such that even the exact black hole solutions can be found.       
\end{abstract}


\maketitle

%%%%%%%%%%%%%%%%%%%%%%%%%%%%%%%%%%%%%%%%%%%%%%%%%%%%%%%%%%%%%%%%%%%%%%%%%%%%%%%%%%%%%%%
\section{Introduction}
K-essence is a certain class of scalar field models with nonstandard kinetic terms.  The name was first coined in a paradigm for inflation  by Armendariz-Picon et al \cite{picon:1999, garriga:1999}. Subsequently, it was found that k-essence can present interesting models for dark energy \cite{chiba:2000,picon:2000,picon:2001,chiba:2002,chimento:2004,chimento:2003,malquarti:2003}. Right after these studies,  Scherrer shows that K-essence can also serve as
the unified model for dark matter and dark energy \cite{robert:2004}. Afterwards, researches on the K-essence theories spring up. 

Specifically, Chimonto and Lazkoz make an analysis on the atypical K-essence cosmologies \cite{chimento:2005}. Abramo et al conduct the research on the stability problem of phantom k-essence theories \cite{abramo:2006}. Bonvin et al \cite{bonvin:2006} demonstrate that if one uses K-essence to solve the coincidence problem and play the role of dark energy in the Universe, the fluctuations of the field have to propagate superluminally at some stage. This implies that successful K-essence models violate causality. Therefore, K-essence cannot arise as a low energy effective field theory of a causal, consistent high energy theory. Bamba \cite{bamba:2012} et al carry out a thorough investigation on the cosmological perturbations of K-essence.  The causality and superluminal behavior K-essence theories  are investigated by Bruneton  \cite{jean:2007}.  The slow-roll conditions for thawing K-essence are derived by Chiba et al with a separable Lagrangian K-essence \cite{chiba:2009}. The K-essence inflation model in the framework of $f(R)$ gravity is studied by Nojiri et al \cite{nojiri:2019}. The phenomenological implications of the constant-roll condition
for the K-essence inflation is investigated by Odintsov and  
Oikonomou \cite{odintsov:2020}.  The late-time behavior of a pure K-Essence in $f(R)$ gravity is studied by Odintsov et al \cite{odintsov:2020pdu}. 
Considering the unification between the swampland criteria and the attractor, Herrera studies the reconstruction of the K-essence inflationary universe \cite{herrera:2020}. Gialamasa and Lahanasa show that the reheating in $R^2$ Palatini gravity in the Einstein frame, shares common features with K-essence models of inflation \cite{gia:2020}. The inflationary phenomenology of a K-essence inflation in the framwork of Einstein-Gauss-Bonnet gravity is conducted by Odintsov et al \cite{odintsov:2021}. Furthermore, K-essence can yield a complete picture of the evolution of the Universe, starting from the early inflation, the subsequent dark matter domination, and to the late time cosmic acceleration \cite{bose:2009a,bose:2009b}. Interestingly, Tian and Zhu's research  \cite{tian:2021} reveals that K-essence framework is also conducive  to the  Hubble tension  problem \cite{riess:2021,freedman:2021}.\footnote{But Lee et al \cite{lee:2022} prove that the EOS (equation of state) of dark energy and the present-day Hubble parameter are anti-correlated in K-essence theories. So K-essence as latetime dark energy may make the Hubble tension even worse. The reason is that if the K-eesence scalar does not rest at the minimum of the potential, the present-day Hubble parameter would be decreased.}

Very recently, Hung and Miao \cite{hung:2023} investigate 
whether a quantum-induced K-essence can provide efficient reheating without affecting the observational constraints on primordial inflation. 
Kehayias and  Scherrer find a new generic evolution for K-essence dark energy with the equation of state of roughly cosmological constant \cite{kehayias:2019}.  
Sarkar et al \cite{sarkar:2023} show that the accretion of k-essence dark energy can lead to a faster circularising of orbits for binary black holes. 
Lara et al \cite{lara:2022} investigate in depth the relation between K-essence, well-posedness of the Cauchy problem and the UV completions.
Huang  \cite{huang:2021} constructs the thawing K-essence models by generating Taylor expansion coefficients of Lagrangian functions from random matrices.
Chatterjee et al make a study on the dynamical stability problem for K-essence interacting nonminimally with a perfect fluid \cite{chatt:2021}.
Applying the ``$2+1+1$'' decomposition of spacetime, Nagy et al conduct the research of  spherically symmetric, static black holes with scalar hair, and naked
singularities in nonminimally coupled K-essence \cite{nagy:2021}. Barvinsky et al find that the generalized unimodular gravity can be transformed into  a new form of K-essence \cite{barvinsky:2021}.  Finally, the K-essence model in which K-essence can be responsible for both primordial
inflation and the present observed acceleration, while also admitting a nonsingular de Sitter beginning of the Universe is achieved by Ferreira et al \cite{fer:2024}.

Up to this point, what we need to emphasize is that all the above researches on K-essence are actually one-field K-essence. In specific, the Lagrangian of K-essence is the function of single field $\phi$ and its derivatives $\partial_{\mu}\phi$. However, in this paper, we shall consider two-field K-essence among which there are two scalar fields $\phi$ and $\psi$ together with their derivatives. The novelty of this research is that the Lagrangian functions are significantly different from the previous ones.  Exactly, they are constructed from the quotient form``potenials'', $x/y$ with $x$ and $y$ the kinetic term of the fields.  To our knowledge,  one has never make an investigation on this sort of K-essence. On the other hand, what we consider is a pure K-essence which makes the calculations considerable simple. In fact, we can get the energy density, the pressure and the equation of state straightforward because the equations of motion of K-essence can be exactly solved without the help of dynamical analysis. 
Although it seems our two-field pure K-essence can be transformed to the multi-scalar tensor theories, the equivalence is merely mathematically, not physically at least from the standpoint of speed of perturbations. The reason is that the resulting metric of field-space is not the function of dynamical field, but the auxiliary scalar fields. It is not the case for conventional multi-scalar tensor theories \cite{langlois:2008}.           

 The paper is organized as follows. In Section II, we derive the Einstein equations and the equations of motion for the two-field pure K-essence. In Section III, we study the cosmic  evolution of the K-essence in the background of Friedmann-Robertson-Walker Universe. In the first place, the Lagrangian functions of K-essence for some conventional EOS are reconstructed. Then motivated by the Lagrangian of K-essence form for the conventional EOS, we explore the so-called novel K-essence Lagrangian which has the quotient ``potentials''. In this way, we obtain several unconventional EOSs (equation-of-states) of K-essence. In Section IV, we transform the two-field pure K-essence to the multi-field form with resort to auxiliary scalar fields.  We point out that the two frameworks are merely equivalent to each other mathematically, not physically, at least from the point of view of speed of perturbations. In Section V, as the demonstration again for the simplicity of novel two-field pure K-essence, we give a black hole solution.  Finally, Section VI gives the conclusion and discussion. Throughout this paper, we adopt the
system of units in which $G=c=\hbar=1$ and the metric signature
$(-, +, +, +)$.

\section{equations of motion}
We consider the action   
\begin{equation}
S=\int{d^4x}\sqrt{-g}\left[\frac{R}{16\pi}+F+L_{m}\right]\;,
\end{equation}
where 
\begin{equation}
F=F\left(x,y\right)\;,
\end{equation}
is the Lagrangian of two-field pure K-essence.  Following Ref.~\cite{robert:2004}  we call it pure K-essence because it does not depend explicitly on the two scalar fields,  $\phi$ and $\psi$.  The Lagrangian is only the function of terms, $x$ and $y$.  Here $x$ and $y$ are the kinetic energy of scalar fields $\phi$ and $\psi$, respectively.  They are defined by 
\begin{equation}
x\equiv -\frac{1}{2}\nabla_{\mu}\phi\nabla^{\mu}\phi\;,\ \ \ y\equiv -\frac{1}{2}\nabla_{\mu}\psi\nabla^{\mu}\psi\;,
\end{equation}
and they are positive with the signature of $(-, +, +, +)$ in the Friedamnn-Roberstson-Walker Universe.   
$R$ is the Ricci scalar and $L_{m}$ is the Lagrangian for matters.  Make variation of the action with respect to the metric tensor, we obtain the Einstein equations
\begin{equation}
G_{\mu\nu}=8\pi\left(T^{k}_{\mu\nu}+T^{m}_{\mu\nu}\right)\;,
\end{equation}
where $T^{m}_{\mu\nu}$ and $T^{k}_{\mu\nu}$  represent  the energy-momentum of matters and K-essence, respectively. The expression of   $T^{k}_{\mu\nu}$  is found to be 
\begin{equation}
T^{k}_{\mu\nu}=F_{,x}\nabla_{\mu}\phi\nabla_{\nu}{\phi}+F_{,y}\nabla_{\mu}\psi\nabla_{\nu}{\psi}+g_{\mu\nu}F\;,
\end{equation}
with the comma denotes the derivative with respect to $x$ and $y$, respectively.  When $F=constant$, we obtain the energy-momentum tensor for the Einstein cosmological constant. When $F=-2x$ or $F=-2y$ ,  we obtain  the energy-momentum tensor for stiff matter. Finally, the variation of the action with respect to $\phi$ and $\psi$ give their equations of motion as follows

\begin{equation}
\nabla_{\mu}\left[F_{,x}\nabla^{\mu}\phi\right]=0\;,\ \ \ \nabla_{\mu}\left[F_{,y}\nabla^{\mu}\psi\right]=0\;.
\end{equation}

\section{cosmic evolution}
In this section, we explore the cosmic evolution of the two-field pure K-essence in the background of Friedmann-Robertson-Walker Universe. To this end, we should first derive the cosmological equations of motion. 
\subsection{cosmological equations of motion}

In the background of Friedmann-Robertson-Walker Universe with the line element 
\begin{equation}
ds^2=-dt^2+a\left(t\right)^2\left(dr^2+r^2d\theta^2+r^2\sin^2\theta d\varphi^2\right)\;,
\end{equation}
where $a(t)$ is the scale factor of the Universe. The Einstein equations appear as the Friedmann equation and the acceleration equation 
\begin{equation}\label{fr}
3H^2=8\pi\left(\frac{\rho_{m0}}{a^3}+\frac{\rho_{r0}}{a^4}+\rho_K\right)\;,
\end{equation}

\begin{equation}\label{acc}
2\dot{H}+3H^2=-8\pi\left(\frac{\rho_{r0}}{3a^4}+p_K\right)\;.
\end{equation}
Here the energy density and pressure of K-essence are 
\begin{equation}\label{density}
\rho_K=-2xF_{,x}-2yF_{,y}+F\;,\ \ \ \ p_K=-F\;.
\end{equation}
 $H$ is the Hubble parameter and the dot denotes the derivative with respect to the cosmic time $t$. $\rho_{m0}$ and $\rho_{r0}$ are the present-day energy density for dark matter and radiation. The equations of motion for the two scalar fields are 
\begin{equation}\label{phipsi}
\left({a^3F_{,x}}\dot{\phi}\right)^{\cdot}=0\;,\ \ \ \ \  \ \left({a^3F_{,y}}\dot{\psi}\right)^{\cdot}=0\;.
\end{equation}
Eqs.~(\ref{fr},\ref{acc},\ref{density},\ref{phipsi}) constitute the cosmological equations of motion. Observing the expressions of density and pressure, we find they are the functions of Lagrangian $F$. So once the equation of state $p_{K}=p_{K}(\rho_K)$ for K-essence is given, we can obtain the expression of $F$ by solving the partial differential equations. In the next subsection, we shall make an investigation on three conventional EOSs by this way.   

\subsection{Three conventional equation-of-states}
The so-called three conventional EOSs are for the Einstein cosmological constant, the constant EOS and the generalized Chaplygin gas. 

\subsubsection{dynamical cosmological constant}

If we assume the EOS for K-essence is that for cosmological constant, taking into account  Eqs.~(\ref{density})  and solving the differential equation 

\begin{equation}
\omega\equiv \frac{p_K}{\rho_K}=-1\;,
\end{equation}
we find 
\begin{equation}\label{cc}
F=f\left(\frac{x}{y}\right)\;.
\end{equation}
Here and thereafter $f\left(\frac{x}{y}\right)$ stands for the  arbitrary function of $x/y$. It is interesting that in this case, although the energy density and pressure of K-essence are constants, the scalar fields $\phi$ and $\psi$ are dynamic: 
\begin{equation}
\dot{\phi}\sim a^3\;, \ \ \ \dot{\psi}\sim a^3\;, 
\end{equation}
for example, when
\begin{equation}\label{ccexample}
F=s\frac{x}{y}\;,
\end{equation}
with $s$ a constant. This  has been pointed out by us in Ref.~\cite{gao:2022}. It is then straightforward to obtain the Schwarzschild-de Sitter solution  
\begin{eqnarray}
ds^2&=&-\left(1-\frac{2M}{r}-\frac{sc_2^2r^2}{6c_1^2}\right)dt^2+r^2d\theta^2+r^2\sin^2\theta d\varphi^2\nonumber\\&&+\left(1-\frac{2M}{r}-\frac{sc_2^2r^2}{6c_1^2}\right)^{-1}dr^2\;,
\end{eqnarray}
in this case. Here $c_1$ and $c_2$ are two integration constants and $M$ is the mass of black hole. The expressions of two scalar fields are 
\begin{eqnarray}
\phi=\frac{2sc_2^2r^3}{3c_1^3}\;,\ \ \ \ \psi=-\frac{2sc_2r^3}{3c_1^2}\;.
\end{eqnarray}
Then the  energy density and pressure in this static spacetime which are the same as in the Friedmann-Robertson-Walker spacetime,  remain constants.  We emphasize that this is a coincidence. In general, for example in the background of  anisotropic and inhomogenious spacetime, the model of Eq.~(\ref{cc})  or Eq.~(\ref{ccexample}) would significantly different from the Einstein cosmological constant because their energy-momentum tensors are obviously different.

\subsubsection{constant equation of state}

If we assume the equation of state for K-essence is a constant $\omega$, taking into account  Eqs.~(\ref{density})  and  solving the 
equation 

\begin{equation}
\omega=\frac{p_K}{\rho_K}\;,
\end{equation}
we find 
\begin{equation}\label{lag}
F=f\left(\frac{x}{y}\right)x^{\frac{1+\omega}{2\omega}}\;.
\end{equation}
 In this case, the energy density and pressure of K-essence are 
\begin{equation}
\rho_K=\frac{\rho_{K0}}{a^{3\left(\omega+1\right)}}\;,\ \ \ \ \ p_K=\omega \rho_K\;.
\end{equation}
It seems the Lagrangian Eq.~(\ref{lag}) is divergent in the limit of  $\omega=0$ (for dark matter). This is an illusion; if we  parameterize the Lagrangian as below, 

\begin{equation}
F=\frac{x^{\frac{1}{2\omega}}}{{e^{\frac{1}{\omega^2}}}}\sqrt{y}\;.
\end{equation}
there is no divergence in the expression when $\omega=0$.

\subsubsection{Generalized Chaplygin gas}

If we assume the equation of state for K-essence is that for generalized Chaplygin gas \cite{pad:2003},  taking into account  Eqs.~(\ref{density})  and solving the 
equation 

\begin{equation}
\rho_{K}=\frac{A}{\left(-p_K\right)^{n-1}}\;,
\end{equation}
we obtain
\begin{equation}\label{chap}
F=\left[A+x^{\frac{n}{2}}f\left(\frac{x}{y}\right)\right]^{\frac{1}{n}}\;,
\end{equation}
where $A$ and $n$ are positive constants. Observations constrain $n>1$  in order to ensure the condition of stability and the condition of causality. 

Up to this point, we conclude that the two-field pure K-essence is very powerful since it can play the role of Einstein cosmological constant, the fluid with a constant EOS and the generalized Chaplygin gas in the back ground of Friedmann-Robertson-Walker Universe. Of course, in the background of other spacetime, the behavior of two-field pure K-essence is certainly different from the others  because they have different energy-momentum tensors.  Observing the Lagrangians Eq.~(\ref{cc}), Eq.~(\ref{lag}) and Eq.~(\ref{chap}), we find they are all the functions of quotient $x/y$. This motivates us, in the next section, to make an investigation on the cosmic evolution of K-essence by starting from a given expression of  $F$ and several unconventional EOSs are found.       
\subsection{four unconventional EOSs}
In this section, we shall first assume the expression of Lagrangian function $F$  and then explore its cosmic evolution. By this way, some novel EOSs are found. 

\subsubsection{The case of $F=2y+s{\frac{\sqrt{x}}{y^{2n}}}$}

In the first place, we consider the Lagrangian function
\begin{equation}
F=2y+s{\frac{\sqrt{x}}{y^{2n}}}\;,
\end{equation}
where $s$ and $n$ are constants. We note that the sign before the kinetic term is positive. This indicates $\phi$ is a phantom scalar field.  Then the cosmological equations of motion, Eqs.~(\ref{density},\ref{phipsi}) give the density and pressure of K-essence as follows

\begin{equation}
\rho_K=b_0a^{\frac{3\left(1-4n\right)}{4n}}+b_1 a^{\frac{3}{2n}}\;,
\end{equation}
\begin{equation}
p_K=-\frac{b_0}{4n}a^{\frac{3\left(1-4n\right)}{4n}}-\frac{b_1\left(2n+1\right)}{2n} a^{\frac{3}{2n}}\;,
\end{equation}
where $b_0$ and $b_1$ are positive integration constants. The equation of state is 

\begin{equation}
\omega=-\frac{\frac{b_0}{4n}a^{\frac{3\left(1-4n\right)}{4n}}+\frac{b_1\left(2n+1\right)}{2n} a^{\frac{3}{2n}}}{b_0a^{\frac{3\left(1-4n\right)}{4n}}+b_1 a^{\frac{3}{2n}}}\;.
\end{equation}
It is interesting that for large $n$, we have the energy density
\begin{equation}
\rho_K=b_0a^{-3}+b_1\;,
\end{equation}
which is the sum of dark matter and cosmological constant. Thus, when $n$ is sufficiently large, the K-essence unifies the cold dark matter and the Einstein cosmological constant. When $n=1/4$, we have 
\begin{equation}
\rho_K=b_0+b_1 a^6\;.
\end{equation}
It consists of two terms among them one is the Einstein cosmological constant and the other is a phantom.  In this case, the K-essence can play the role of dark energy for present-day cosmic acceleration .

\begin{figure}[htbp]
	\centering
	\includegraphics[width=8cm,height=6cm]{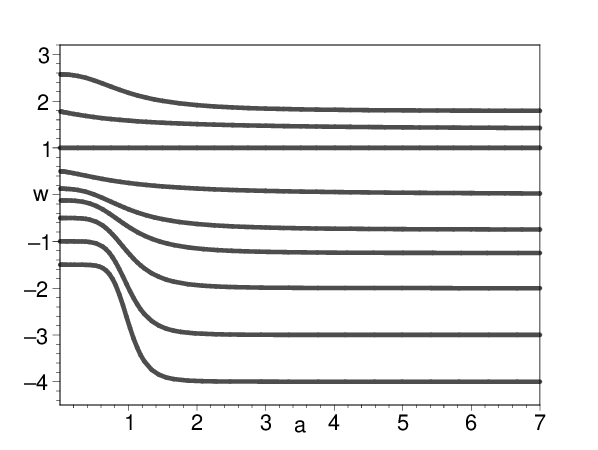}
	\caption{The evolution of EOS  for K-essence with respect to the scale factor when $n=0.17,\ 0.25,\ 0.5,\ 2$ from down to top, respectively for the lower half of the graph.  For the upper half of the graph, we put  $n=-0.14,\ -0.18,\ -0.25,\ -0.5,\ -2$ from up to down. We assume $b_0=b_1=1$.} 
	\label{eos1-1.eps}
\end{figure}

In Fig.~(\ref{eos1-1.eps}) we plot the equation of state $\omega$  with respect to the scale factor. In order that the energy density is always positive, we assume the two integration constants $b_0>0$ and $b_1>0$. The figure shows that when $n=-1/4$, the K-essence behaves as the stiff matter which has the equation of state $+1$. However, it is amazing that its Lagrangian function $F=2y+s\sqrt{xy}$ is obviously different from that for conventional stiff matter $F=-2y$. If $n=1/4$, the K-essence can play the role of cosmic dark energy as mentioned above. On the other hand, if $|n|\rightarrow\infty$, the K-essence unifies the Einstein cosmological constant and the cold dark matter model.  Finally, when $n\approx{0}$, the equation of state can be very large such that $w\rightarrow {+\infty}$. Now the K-essence becomes the stiffest matter in nature.  

\subsubsection{The case of $F=s_1 \sqrt{x}+s_2\sqrt{y}+s\left(\frac{{x}}{y}\right)^{\frac{n}{2}}$}
Secondly, we consider the Lagrangian function 
\begin{equation}
F=s_1 \sqrt{x}+s_2\sqrt{y}+s\left(\frac{{x}}{y}\right)^{\frac{n}{2}}\;,
\end{equation}
where $s_1,\  s_2,\ s$ and $n$ are constants. We recognize that $s_1$ and $s_2$ terms are actually the kinetic terms of Cuscuton action \cite{afshordi:2007}.  The cosmological equations of motion, Eqs.~(\ref{density},\ref{phipsi}) give the energy density 

\begin{equation}
\rho_K=-\frac{s\left(c_2-s_2a^3\right)^n}{\left(c_1-s_1a^3\right)^{n}}\;,
\end{equation}
where $c_1$ and $c_2$ are integration constants. It is straightforward to obtain the pressure $p_K$ and equation of state $\omega$ for K-Essence by substituting the energy density into the energy conservation equation 
\begin{equation}\label{ece}
\frac{d\rho_K}{dt}+3H\left(\rho_K+p_K\right)=0\;.
\end{equation}
In order that the energy density is always positive, we let 
\begin{equation}
s<0\;,\ \ s_1<0\;,\ \ \ s_2<0\;,\ \ c_1>0\;,\ \ c_2>0\;.
\end{equation}
Of course, there are other options available dependant on the values of $n$. 
\begin{figure}[htbp]
	\centering
	\includegraphics[width=8cm,height=6cm]{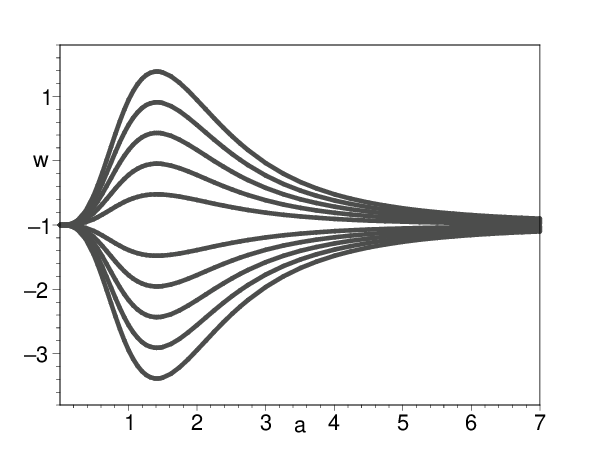}
	\caption{The evolution of EOS for K-essence with respect to the scale factor when $n=-5\;,-4\;,-3\;,-2\;,-1\;, 1\;,2\;, 3\;,4\;, 5$ from down to top, respectively.  We have put $s_1=s_2=-1,\ c_1=1,\ c_2=8$.}\label{eos2-1.eps}
\end{figure}

\begin{figure}[htbp]
	\centering
	\includegraphics[width=8cm,height=6cm]{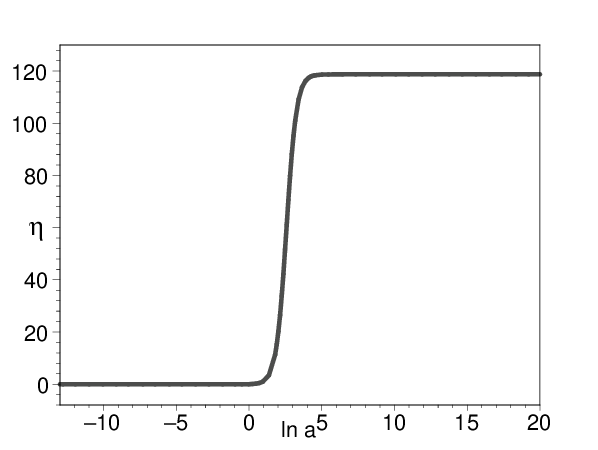}
	\caption{The evolution of for the Log-ratio $\eta$ of K-essence density to its present-day value with respect to the scale factor. We have put $s_1=-2\cdot 10^{-3},\ s_2=-10^{-3},\ c_1=10,\ c_2=1,\ n=170$.}\label{eos2-2.eps}
\end{figure}

In Fig.~(\ref{eos2-1.eps}) we plot the EOS $\omega$ with respect to the scale factor. It shows that the EOS starts from $-1$ at very early universe and then decreases to negative. Subsequently it approaches $-1$ in the distant future. This shows the energy density emerges as nearly a constant at very early universe. Then it gets a remarkable growth. Eventually, the energy density climbs to a very huge constant again by the order of $120$ which is the order Planck energy density. To show this point, we plot the evolution for the Log-ratio 

\begin{equation}
\eta=\log{\frac{\rho_K}{\rho_{K0}}}. 
\end{equation}
of K-essence density $\rho_K$ to its present-day value $\rho_{K0}$ in Fig.~(\ref{eos2-2.eps}). Thus we conclude that the fate of our Universe is to undergo an inflation a third time.     

\begin{figure}[htbp]
	\centering
	\includegraphics[width=8cm,height=6cm]{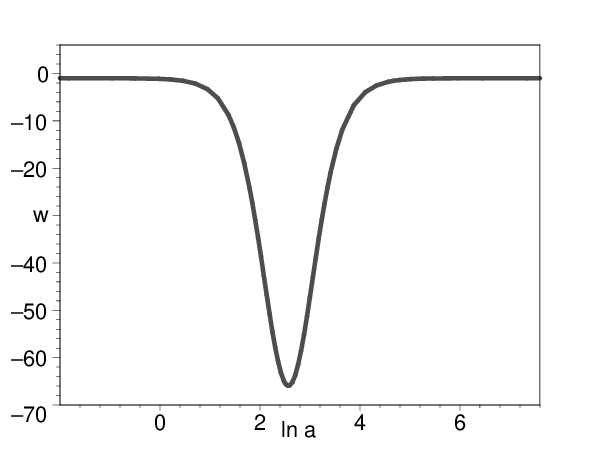}
	\caption{The evolution of EOS for K-essence with respect to the natural logarithm of scale factor. We have put $s_1=-2\cdot 10^{-3},\ s_2=-10^{-3},\ c_1=10,\ c_2=1,\ n=170$.}\label{eos2-3.eps}
\end{figure}

\begin{figure}[htbp]
	\centering
	\includegraphics[width=8cm,height=6cm]{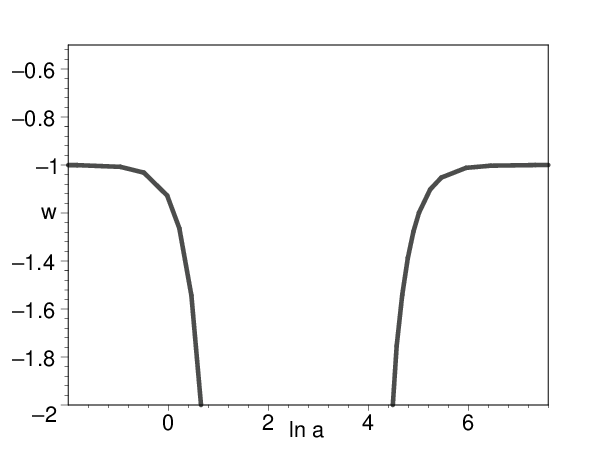}
	\caption{The evolution of EOS for K-essence being a part of Fig.~(\ref{eos2-3.eps}), with respect to the natural logarithm of scale factor.}\label{eos2-4.eps}
\end{figure}

In Fig.~(\ref{eos2-3.eps}) and Fig.~(\ref{eos2-4.eps}) we plot the evolution of EOS for K-essence with respect to the natural logarithm of scale factor.
The present-day EOS is slightly less than $-1$ which is consistent with the astronomical observations.

\begin{figure}[htbp]
	\centering
	\includegraphics[width=8cm,height=6cm]{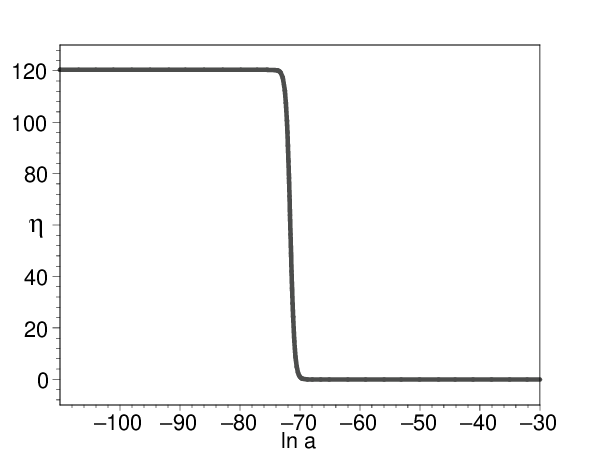}
	\caption{The evolution of for the Log-ratio $\eta$ of K-essence density to its present-day value with respect to the natural logarithm of scale factor. We have put $s_1=-4,\ s_2=-1,\ c_1=10^{-93},\ c_2=10^{-93},\ n=200$.}\label{eos2-5.eps}
\end{figure}

\begin{figure}[htbp]
	\centering
	\includegraphics[width=8cm,height=6cm]{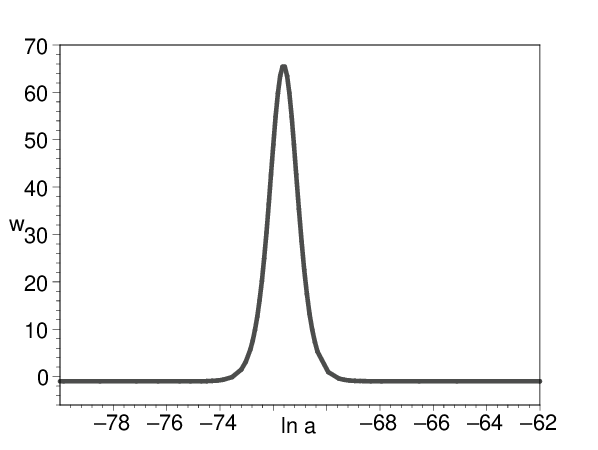}
	\caption{The evolution of EOS for K-essence with respect to the natural logarithm of scale factor. We have put $s_1=-4,\ s_2=-1,\ c_1=10^{-93},\ c_2=10^{-93},\ n=200$.}\label{eos2-6.eps}
\end{figure}

\begin{figure}[htbp]
	\centering
	\includegraphics[width=8cm,height=6cm]{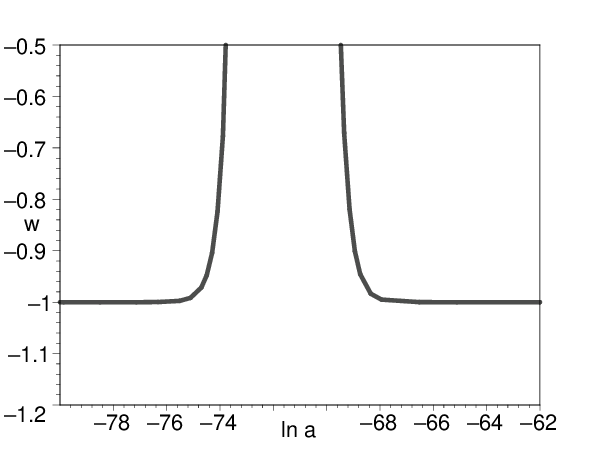}
	\caption{The evolution of EOS for K-essence being a part of Fig.~(\ref{eos2-6.eps}), with respect to the natural logarithm of scale factor.}\label{eos2-7.eps}
\end{figure}

On the other hand, Fig.~(\ref{eos2-1.eps}) shows that the EOS can also  start from  $-1$ at very early universe and then increases to positive, its curves bulging and distended. In the end, it arrives at $-1$. This shows the energy density is nearly a constant at very early universe and then it decreases significantly. Finally, the energy density dwells on a very small constant. This indicates that the K-essence can unify the early inflation field and the later dark energy. In order to make this clear, we plot the evolution for the Log-ratio $\eta$ of K-essence density to its present-day value, the equation of state of K-essence with respect to the natural logarithm  of scale factor in Fig.~(\ref{eos2-5.eps}), Fig.~(\ref{eos2-6.eps}) and Fig.~(\ref{eos2-7.eps}). 

The Hubble radius of the present-day Universe is the order of $10^{25}$ meter. Inflation is generally assumed starting from the Planck length, $10^{-35}$ meter. So on the whole, the scale factor expands by the order of $10^{60}$ or $140$ e-folds. In order to solve the horizon problem, $60$ e-folds more or less from inflation is needed. Without the loss of generality, we assume inflation produces $70$ e-folds. Therefore, inflation starts from $a\sim e^{-140}$ to $a\sim e^{-70}$.  Then from  $a\sim e^{-70}$ to $a\sim e^{0}$, the universe goes through three stages, namely, the radiation-dominated, matter-dominated and matter plus dark-energy dominated stages, respectively,

Fig.~(\ref{eos2-5.eps}) tells us the early density is the order of $120$ compared to the present-day cosmological constant. So the cosmological constant problem disappears. Fig.~(\ref{eos2-6.eps}) shows the EOS can be very large which indicates the K-essence energy density undergoes a precipitous decline at this instant. Fig.~(\ref{eos2-7.eps}) clearly shows the EOS has $-1$ as its starting point and finally consequently result. 

In the next subsection, we shall present the other example which can unify the early inflation field and  the latter cosmic dark energy.

\subsubsection{The case of $F=s_1 \sqrt{x}+s_2\sqrt{y}+se^{n\left(\frac{{x}}{y}\right)^{\frac{m}{2}}}$}
\begin{figure}[htbp]
	\centering
	\includegraphics[width=8cm,height=6cm]{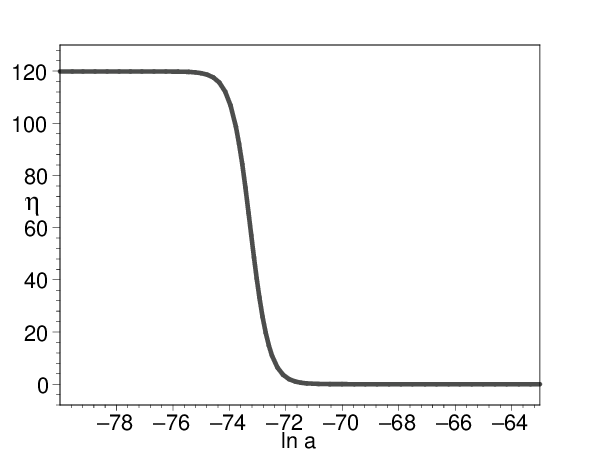}
	\caption{The evolution of for the Log-ratio $\eta$ of K-essence density to its present-day value with respect to the natural logarithm of scale factor. We have put $s_1=-277,\ s_2=-1,\ c_1=10^{-93},\ c_2=10^{-93},\ n=-277$.}\label{unifydensity.eps}
\end{figure}

\begin{figure}[htbp]
	\centering
	\includegraphics[width=8cm,height=6cm]{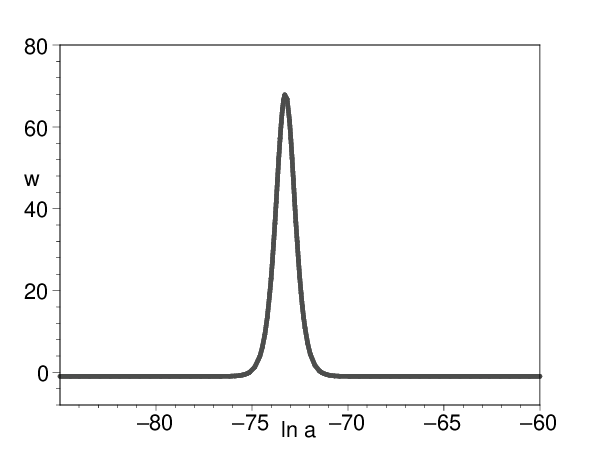}
	\caption{The evolution of EOS for K-essence with respect to the natural logarithm of scale factor. We have put $s_1=-277,\ s_2=-1,\ c_1=10^{-93},\ c_2=10^{-93},\ n=-277$.}\label{unifyeos1.eps}
\end{figure}

\begin{figure}[htbp]
	\centering
	\includegraphics[width=8cm,height=6cm]{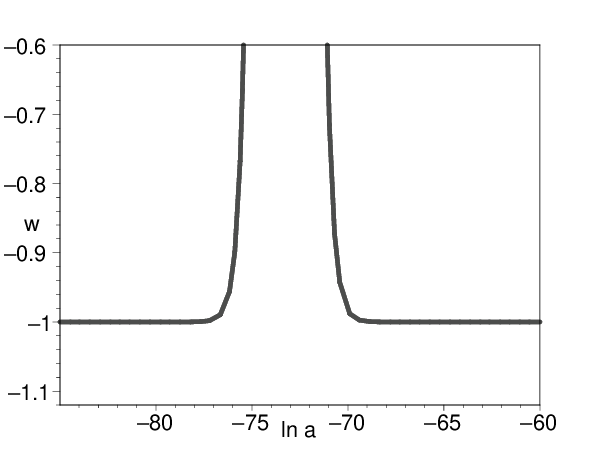}
	\caption{The evolution of EOS for K-essence being a part of Fig.~(\ref{unifyeos1.eps}), with respect to the natural logarithm of scale factor.}\label{unifyeos2.eps}
\end{figure}

Thirdly, we consider the Lagrangian 
\begin{equation}\label{infDE}
F=s_1 \sqrt{x}+s_2\sqrt{y}+se^{n\left(\frac{{x}}{y}\right)^{\frac{m}{2}}}\;,
\end{equation}
where $s_1,\ s_2,\ s,\ n$ and $m$ are constants.  The cosmological equations of motion, Eqs.~(\ref{density},\ref{phipsi}) give the energy density 
\begin{equation}
\rho_K=s e^{n\frac{\left(c_{2}-a^{3}s_{2}\right)^{m}}{\left(a^{3}s_{1}-c_{1}\right)^{m}}}\;,
\end{equation}
with $c_1$ and $c_2$ two integration constants. When the scale factor $a$ is sufficiently small, i.e. $a\rightarrow 0$, we obtain a constant  energy density 

\begin{equation}
\rho_{Ke}=s e^{n\frac{c_{2}^m}{\left(-c_{1}\right)^{m}}}\;.
\end{equation}
On the other hand, when the scale factor is sufficient large, we also have a constant energy density 

\begin{equation}
\rho_{Kl}=s e^{n\frac{\left(-s_{2}\right)^{m}}{s_{1}^{m}}}\;.
\end{equation}
The Log-ratio of above constant densities is

\begin{equation}
\eta\equiv\log{\frac{\rho_{Ke}}{\rho_{Kl}}}=\frac{1}{\ln{10}}\left[{n\frac{c_{2}^m}{\left(-c_{1}\right)^{m}}-n\frac{\left(-s_{2}\right)^{m}}{s_{1}^{m}}}\right]\;.
\end{equation}
If we require 

\begin{equation}
n\frac{c_{2}^m}{\left(-c_{1}\right)^{m}}\sim {277}\;,\ \ \  n\frac{\left(-s_{2}\right)^{m}}{s_{1}^{m}}\sim{1}\;,
\end{equation}
we would find 

\begin{equation}
{\eta}\sim{{120}}\;.
\end{equation}
Namely, the constant energy density for inflation is $120$ order of magnitude  larger than the present dark energy density.   To make this clear and for simplicity, we consider the parameter 
\begin{equation}
m=1\;.
\end{equation}
Then we obtain the energy density 
\begin{equation}
\rho_{K}=s e^{-n\frac{a^3s_2-c_2}{a^3s_1-c_1}}\;.
\end{equation}
Given that the energy conservation equation Eq.~(\ref{ece}), the EOS can be obtained. In Fig.~(\ref{unifydensity.eps}), Fig.~(\ref{unifyeos1.eps}) and Fig.~(\ref{unifyeos2.eps}),  we plot the evolution for the Log-ratio $\eta$ of K-essence density to its present-day value, the EOS of K-essence with respect to the natural logarithm of scale factor. To conclude, we find the Lagrangian function Eq.~(\ref{infDE}) can be used as a model of unification of early inflation field and later cosmic dark energy.

\subsubsection{The case of $F=s_1 \sqrt{x}+s_2\sqrt{y}+s\frac{\sqrt{x}}{y^{\frac{n}{2}}}$}
\begin{figure}[htbp]
	\centering
	\includegraphics[width=8cm,height=6cm]{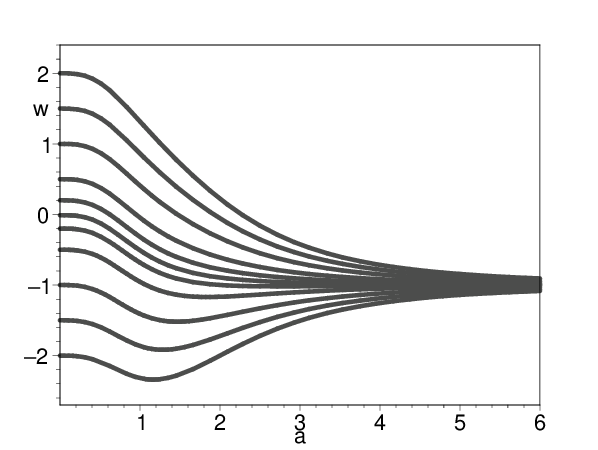}
	\caption{The evolution of EOS  for K-essence with respect to the scale factor when $n=-\frac{1}{2}\;,-\frac{2}{3}\;,-1\;,-2\;,-5$ from top to down for the upper half of the graph.  For the lower half of the graph, we put $n=\frac{1}{2}\;,\frac{2}{3}\;,1\;,5\;,100$ from down to top, respectively. We set  $s_1=-1\;,s_2=1\;,c_2=-1\;,c_1=10$ fro both upper and lower graphs.}\label{eos3-1.eps}
\end{figure}

\begin{figure}[htbp]
	\centering
	\includegraphics[width=8cm,height=6cm]{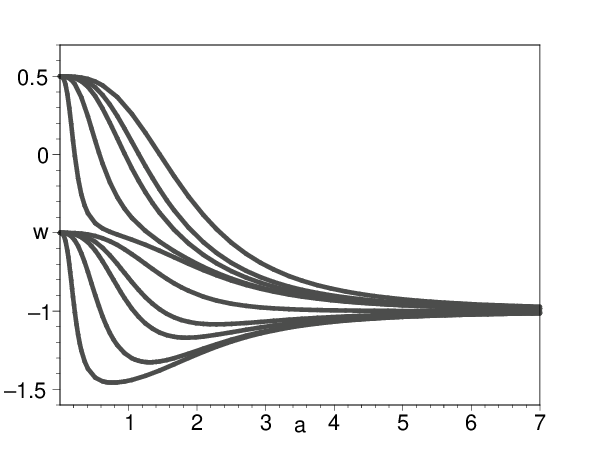}
	\caption{The evolution of EOS  for K-essence with respect to the scale factor when $n=-2$ and $n=2$ for the upper and lower half of the graph, respectively.  We set $s_2=100\;,5\;,1\;,0.5\;,0.25$ from down to top, respectively, for both halves. We also set  $s_1=-1\;,c_2=-1\;,c_1=10$ fro both upper and lower graphs.}\label{eos3-2.eps}
\end{figure}

\begin{figure}[htbp]
	\centering
	\includegraphics[width=8cm,height=6cm]{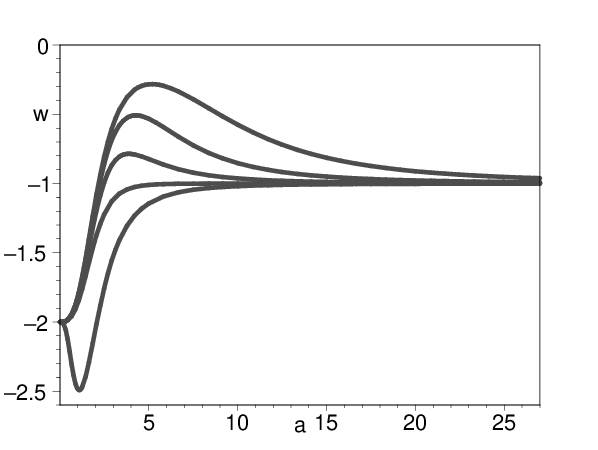}
	\caption{The evolution of EOS  for K-essence with respect to the scale factor when $c_2=-0.5\;,-20\;,-60\;,-200\;,-800$ from  down to top.  We set $n=\frac{1}{2}\;, s_1=-1\;,s_2=1\;,c_1=10$ .}\label{eos3-3.eps}
\end{figure}

\begin{figure}[htbp]
	\centering
	\includegraphics[width=8cm,height=6cm]{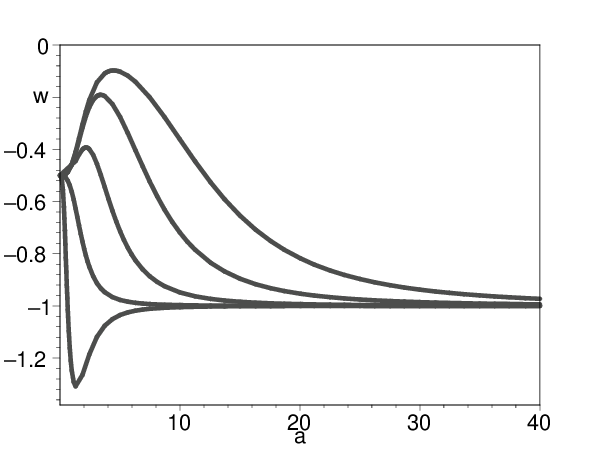}
	\caption{The evolution of EOS  for K-essence with respect to the scale factor when $c_2=-0.25\;,-8\;,-60\;,-400\;,-1800$ from  down to top.  We set $n=2\;, s_1=-1\;,s_2=1\;,c_1=10$ .}\label{eos3-4.eps}
\end{figure}

Finally, we consider the Lagrangian 
\begin{equation}
F=s_1 \sqrt{x}+s_2\sqrt{y}+s\frac{\sqrt{x}}{y^{\frac{n}{2}}}\;,
\end{equation}
where $s_1,\ s_2,\ s$ and $n$ are constants. The cosmological equations of motion, Eqs.~(\ref{density},\ref{phipsi}) give the energy density 

\begin{equation}
\rho_K=\frac{1}{n}{s^{\frac{1}{n}}\left(s_2a^3-c_2\right)a^{-3+\frac{3}{n}}}{\left(c_1-s_1a^3\right)^{-\frac{1}{n}}}\;,
\end{equation}
where $c_1$ and $c_2$ are integration constants. It is straightforward to obtain the pressure $p_K$ and thus equation of state $\omega$ 

\begin{equation}
\omega=-\frac{ns_1s_2a^6-ns_2c_1a^3-s_2c_1a^3+c_1c_2}{n\left(s_2a^3-c_2\right)\left(s_1a^3-c_1\right)}\;,
\end{equation}
for K-Essence by substituting the energy density into the energy conservation equation. In Fig.~(\ref{eos3-1.eps}), Fig.~(\ref{eos3-2.eps}), Fig.~(\ref{eos3-3.eps}) and Fig.~(\ref{eos3-4.eps}), we plot the EOS with respect to the scale factor. Fig.~(\ref{eos3-1.eps}) shows the EOS can be very positive or negative at the early universe. The large positive and negative EOS means the energy density decreases and increases, respectively, very fast. When $a\rightarrow{0}$, we have the EOS 
\begin{equation}\label{earlyw}
\omega=-\frac{1}{n}\;.
\end{equation}
Then when $n\rightarrow\pm\infty $, we have $\omega=0$ as shown in Fig.~(\ref{eos3-1.eps}). 
When $a\rightarrow\infty$, we have 
\begin{equation}\label{laterw}
\omega=-1\;,
\end{equation}
which is also shown in Fig.~(\ref{eos3-1.eps}). Fig.~(\ref{eos3-2.eps}), Fig.~(\ref{eos3-3.eps}) and Fig.~(\ref{eos3-4.eps}) reveal that the EOS can cross the phantom divide either from large to small as shown in Fig.~(\ref{eos3-2.eps})  or from small to large as shown in Fig.~(\ref{eos3-3.eps}). They have one thing in common. Namely, they all approaches $-1$ in distant future.  

\begin{figure}[htbp]
	\centering
	\includegraphics[width=8cm,height=6cm]{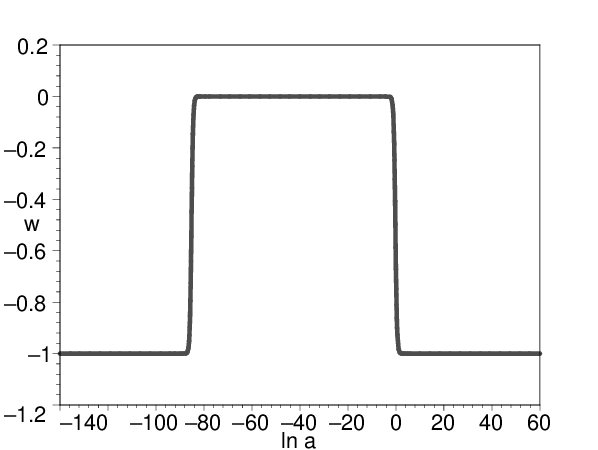}
	\caption{The evolution of EOS  for K-essence with respect to the natural logarithm of scale factor when $c_2=-1\;, s_1=-10^{-14}\;,s_2=2\;,c_1=10^{-125}$ .}\label{tripeunion-2.eps}
\end{figure}

\begin{figure}[htbp]
	\centering
	\includegraphics[width=8cm,height=6cm]{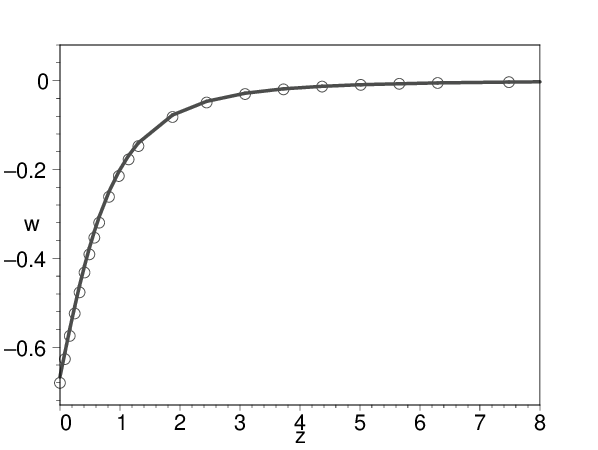}
	\caption{The evolution of EOS  for K-essence with respect to the cosmological redshift when $c_2=-1\;, s_1=-10^{-14}\;,s_2=2\;,c_1=10^{-125}$ . The circled line is for the $\Lambda {CDM}$ paradigm. }\label{tripeunion-3.eps}
\end{figure}

\begin{figure}[htbp]
	\centering
	\includegraphics[width=8cm,height=6cm]{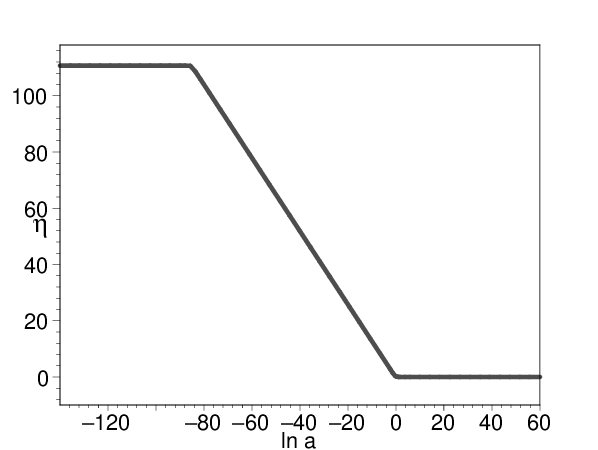}
	\caption{The evolution of for the Log-ratio $\eta$ of K-essence density to its present-day value with respect to the natural logarithm of scale factor when $c_2=-1\;, s_1=-10^{-14}\;,s_2=2\;,c_1=10^{-125}$ .}\label{tripeunion-1.eps}
\end{figure}

It is of particular interest when $n=1$. We find that in this case, the K-essence can unify inflation field, dark matter and dark energy. Eq.~(\ref{earlyw}) and Eq.~(\ref{laterw}) tell us the EOS is $-1$ both at very early universe and later universe which is the desired feature to be the candidate of inflation field and dark energy. In Fig.~(\ref{tripeunion-2.eps}) we plot the EOS of K-essence  with respect to the natural logarithm of scale factor when $n=1$.  The figure shows inflation starts from $a\sim e^{-140}$ to $a\sim e^{-80}$, i.e., $60$ e-folds are generated by inflation. Then from $a\sim e^{-80}$ to $a\sim e^{-0}$, K-essence plays the role of dark matter with EOS of $0$. Finally, at the vicinity of $a^{0}$, namely, the present-day Universe, K-essence plays the role of both dark matter and dark energy. To illustrate this point, we plot the evolution of EOS with respect to cosmic redshift in Fig.~(\ref{tripeunion-3.eps}). The circled line is for $\Lambda CDM$ paradigm. As we can see, the two curves are almost overlap with each other, suggesting that the unification model holds here. In Fig.~(\ref{tripeunion-1.eps}), we plot the evolution for the Log-ratio $\eta$ of K-essence density to its present-day value with respect to the natural logarithm of scale factor. There is a slope between two platforms. The slope denotes the decrease of dark matter with the expansion of the Universe. 

To summarize at this stage, the two-field pure K-essence is a potential candidate for the unification of inflation field, dark matter and dark energy. In the next section, we turn to the issue of speed of perturbations for K-essence.   

\section{on the speed of perturbations}
All the Lagrangian functions studied in the previous section are covered by
\begin{eqnarray}\label{frame1}
F&=&2y+s_1\sqrt{x}+s_2\sqrt{y}+s\frac{x^{\alpha}}{y^{\beta}}+se^{n\left(\frac{x}{y}\right)^{\frac{m}{2}}}\;.
\end{eqnarray}
The first three terms are the kinetic terms and the last two can be understood as ``potential'' terms. The ``potential'' terms are novel in that they consist of quotient expression of kinetic energies.

 The above unified Lagrangian can be written as 
\begin{eqnarray}\label{frame2}
F&=&\left(\phi_1+s\alpha \phi_3 +\phi_5\right)x+\left(2+\phi_2-s\beta\phi_4 -\phi_6\right) y\nonumber\\&&+s\left(1-\alpha+\beta\right){\phi_3^{-\frac{\alpha}{1-\alpha+\beta}}}{\phi_4^{\frac{\beta}{1-\alpha+\beta}}}\nonumber\\&&+\frac{1}{4}\frac{s_1^2}{\phi_1}+\frac{1}{4}\frac{s_2^2}{\phi_2}+se^{n\left(\frac{\phi_6}{\phi_5}\right)^{\frac{m}{2}}}\;,
\end{eqnarray}
by the aid of six auxiliary scalar fields, $\phi_i$ with $i=1-6$. Now the Lagrangian presents itself as a scalar theory with eight scalar fields. Namely, two dynamical scalars $\phi$, $\psi$ which are equipped with the kinetic terms plus six auxiliary scalars $\phi_i$. The auxiliary scalars make a contribution of a scalar potential, $V(\phi_i)$. 
\begin{eqnarray}
V&=&s\left(1-\alpha+\beta\right){\phi_3^{-\frac{\alpha}{1-\alpha+\beta}}}{\phi_4^{\frac{\beta}{1-\alpha+\beta}}}\nonumber\\&&+\frac{1}{4}\frac{s_1^2}{\phi_1}+\frac{1}{4}\frac{s_2^2}{\phi_2}+se^{n\left(\frac{\phi_6}{\phi_5}\right)^{\frac{m}{2}}}\;,
\end{eqnarray}
We recognize immediately that it seems belonging to the generalized multi-field theories \cite{langlois:2008} where
the linear perturbations are already completed.  Then we conclude from Ref. \cite{langlois:2008}  that both the entropy modes
and adiabatic mode propagate with the speed of light in the framework of Lagrangian function Eq.~(\ref{frame2}). Now there is an important issue that we need to direct our attention to. The issue is that the speed of perturbations depend on the framework or gauge for the theory. Namely, the speed of perturbations in the framework of Eq.~(\ref{frame1})
is different from that in the framework of Eq.~(\ref{frame2}). We can understand this point from a simple example. For simplicity, we consider one-field pure K-essence with the Lagrangian 
\begin{eqnarray}\label{example}
F&=&x^2\;,
\end{eqnarray}
It is well-known that the speed of perturbations in this framework is \cite{garriga:1999}
\begin{eqnarray}
c_s^2&\equiv&\frac{F_{,x}}{F_{,x}+2xF_{,xx}}=\frac{1}{2}\;.
\end{eqnarray}
However, the Lagrangian function Eq.~(\ref{example}) can be transformed to be 
\begin{eqnarray}\label{example}
F&=&\phi_1 x-\frac{\phi_1^2}{4}\;,
\end{eqnarray}
by introducing the auxiliary scalar field $\phi_1$. Then the speed of perturbations in this framework is \cite{langlois:2008}
\begin{eqnarray}
c_s^2&\equiv&\frac{F_{,x}}{F_{,x}+2xF_{,xx}}=1\;.
\end{eqnarray}
Therefore, from the perspective of speed of perturbations, the two frameworks are merely equivalent to each other mathematically, not physically. 

\section{black hole solutions}
In the last section, we turn our attention to seek for black hole solutions with two-field pure K-essence. We find it is not a hard work because of the absence of scalar fields in the Lagrangian except for their derivatives.

\subsection{solutions}
The line element of static and spherically symmetric solution for black holes can always be written as  
\begin{equation}
ds^2=-U\left(r\right)dt^2+\frac{N\left(r\right)^2}{U\left(r\right)}dr^2+r^2d\theta^2+r^2\sin^2\theta d\varphi^2\;.
\end{equation}
Since the spacetime is static and spherically symmetric,  the two scalar fields $\phi$ and $\psi$ are the functions of radial coordinate $r$, namely, $\phi=\phi(r)$ and $\psi=\psi(r)$. 
       
At this moment in time, we are interested in the expression of $F$ as follows 
        
\begin{equation}
F=2y+s\sqrt{\frac{x}{y}}\;,
\end{equation}
with $s$ a positive constant.  Then  the Einstein equations are found to be 

\begin{equation}
\frac{U^{'}}{rN^2}-\frac{2UN^{'}}{rN^3}-\frac{1}{r^2}+\frac{U}{r^2N^2}=8\pi\left(\frac{U\psi^{'2}}{N^2}+s\frac{\phi^{'}}{\psi^{'}}\right)\;,
\end{equation}

\begin{equation}
\frac{U^{'}}{rN^2}-\frac{1}{r^2}+\frac{U}{r^2N^2}=8\pi\left(-\frac{U\psi^{'2}}{N^2}+s\frac{\phi^{'}}{\psi^{'}}\right)\;,
\end{equation}

\begin{equation}
\frac{U^{'}}{rN^2}-\frac{UN^{'}}{rN^3}+\frac{U^{''}}{2N^2}-\frac{U^{'}N^{'}}{2N^3}=8\pi\left(\frac{U\psi^{'2}}{N^2}+s\frac{\phi^{'}}{\psi^{'}}\right)\;,
\end{equation}
and the equations of motion for $\phi$ and $\psi$ are 
\begin{equation}
\left(\frac{Nr^2}{\psi^{'}}\right)^{'}=0\;,\ \ \ \ \  \ \left(-\frac{2Ur^2\psi^{'}}{N}+\frac{sr^2N\phi^{'}}{\psi^{'2}}\right)^{'}=0\;.
\end{equation}
Here the apostrophe $'$ stands for the derivative with respect to the coordinate variable $r$.  Solving these equations, we find the solutions

\begin{equation}
\psi=\frac{\sqrt{6}}{12\sqrt{\pi}}arcsinh{\left(\frac{r^6}{a_3^6}\right)}\;,
\end{equation}
\begin{eqnarray}
\phi^{'}&=&-r^5\int\frac{3}{32\pi^{\frac{3}{2}}}\cdot\frac{\sqrt{6}\sqrt{r^6+a_3^6}\left(2a_1r^4+a_2a_3^6\right)}{sa_1r^4\left(r^2+a_3^2\right)^2\left(r^4-r^2a_3^2+a_3^4\right)^2}{dr}\nonumber\\&&+a_4r^5\;, 
\end{eqnarray}

\begin{equation}
N=\frac{a_1}{\sqrt{r^6+a_3^6}}\;,
\end{equation}

\begin{eqnarray}
&&U=\frac{2a_1^2\sqrt{r^6+a_3^6}}{3a_3^9}hypergoem\left[\left(\frac{1}{2},\frac{1}{6}\right),\left(\frac{7}{6}\right),-\frac{r^6}{a_3^6}\right]\nonumber\\&&+\frac{a_1^2}{3a_3^6}+\frac{a_1a_2r^2}{3a_3^6}+\frac{8\pi^{\frac{3}{2}}}{9r}\sqrt{6}sa_1^2a_4\sqrt{r^6+a_3^6}\;.
\end{eqnarray}
Here the parameters $a_i$ with $i=1,2,3,4$ are integration constants. In the limit of $r\rightarrow 0$, we have 
\begin{eqnarray}\label{lim1}
&&U=\frac{a_1^2}{{a_3^6}}+\frac{8\sqrt{6}sa_4a_1^2a_3^3\pi^{\frac{3}{2}}}{9r}\;.
\end{eqnarray}
On the other side, in the limit of $r\rightarrow +\infty$, we know  
\begin{eqnarray}\label{lim2}
&&\frac{2a_1^2\sqrt{r^6+a_3^6}}{3a_3^9}hypergoem\left[\left(\frac{1}{2},\frac{1}{6}\right),\left(\frac{7}{6}\right),-\frac{r^6}{a_3^6}\right]\nonumber\\&&\propto r^2\;,
\end{eqnarray}
thus 
\begin{eqnarray}
U\propto r^2\;,
\end{eqnarray}
when $r\rightarrow +\infty$. 

\subsection{physical significance of the parameters}

By expanding the expression of $U$ into a power series of $r$, we obtain   
\begin{equation}
U=\frac{a_1^2}{{a_3^6}}+\frac{8\sqrt{6}sa_4a_1^2a_3^3\pi^{\frac{3}{2}}}{9r}+\frac{a_1a_2}{3a_3^6}r^2+O(r^5)\;.
\end{equation}
Taking account of Eq.~(\ref{lim1}) and Eq.~(\ref{lim2}), we are aware that we should  compare the metric function  with the Schwarzschild-de Sitter solution. Then the parameters are required to satisfy 
\begin{equation}
\frac{a_1^2}{{a_3^6}}=1\;,\ \ \ \frac{8\sqrt{6}sa_4a_1^2a_3^3\pi^{\frac{3}{2}}}{9}=-2M\;.
\end{equation}
So we obtain 

\begin{equation}
{a_1}={{a_3^3}}\;,\ \ \ a_4=-\frac{3\sqrt{6}M}{8sa_3^9\pi^{\frac{3}{2}}}\;.
\end{equation}
Then the metric functions turn out to be 

\begin{eqnarray}\label{U}
&&U=\frac{2}{3}\frac{\sqrt{r^6+a_3^6}}{a_3^2}hypergoem\left[\left(\frac{1}{2},\frac{1}{6}\right),\left(\frac{7}{6}\right),-\frac{r^6}{a_3^6}\right]\nonumber\\&&+\frac{1}{3}-\frac{2M\sqrt{r^6+a_3^6}}{a_3^3r}+\frac{a_2r^2}{3a_3^3}\;,
\end{eqnarray}
\begin{equation}\label{N}
N=\frac{a_3^3}{\sqrt{r^6+a_3^6}}\;.
\end{equation}
We observe that  in the limit of 

\begin{eqnarray}
&&{r}\ll{a_3}\;,
\end{eqnarray}
we have
\begin{eqnarray}
&&\frac{\sqrt{r^6+a_3^6}}{a_3^2}hypergoem\left[\left(\frac{1}{2},\frac{1}{6}\right),\left(\frac{7}{6}\right),-\frac{r^6}{a_3^6}\right]=1\;.\nonumber
\end{eqnarray}
Then the solution goes back to the Schwarzschild-de Sitter solution for $a_2<0$ and Schwarzschild-anti-de Sitter solution for $a_2>0$.  Here $M$ represents the mass of the black hoe. It is obvious  $a_2$ and $a_3$   have the dimension of mass and they can be understood as the scalar charges of field $\phi$ and $\psi$, respectively.  Similar to the  Schwarzschild-de Sitter solution, the spacetime described by Eq.~(\ref{U}) and Eq.~(\ref{N}) have generally two horizons,i.e, the black hole horizon and the cosmic horizon.  

\section{conclusion and discussion}
In conclusion, we have studied a novel two-field pure K-essence whose Lagrangian is the function of kinetic terms of two scalar fields while independent on the scalar fields. The absence of the scalar fields variables in the Lagrangian makes the equations of motion for scalar fields considerable simple and their analytic solutions can be found. The novelty of this K-essence lies in the ``potential'' which is made up with quotient of kinetic terms of the scalar fields. 

We find that this novel K-essence have many interesting features. They are listed as follows. 

1. The equation of state covers a very large parameter-space, i.e. $-\infty<\omega<+\infty$ which is remarkably different from that of quintessence, i.e. $-1<\omega<+1$.

2. The EOS starts from $-1$, then increases or decreases, and ends in $-1$. Then two fates of the Universe are laid before us. One is a de Sitter universe governed by a very small constant density and the other is an inflationary universe governed by a very large constant density.

3.  The EOS crosses the phantom divide either from positive to negative or from negative to positive. 

4. Since the EOS begins from $-1$, goes through $0$ and ends in $-1$, K-essence unifies the early inflation field,  the later dark energy and the dark matter in between.  

5. The simplicity of the equations of motion enables us to obtain the exact solution of black holes sourced by the K-essence. 

6. The equivalence of this novel K-essence to the multi-field theory is merely from the perspective of mathematics, not physics. The reason is that the metric of field-space is not the function of dynamical scalar fields, but the function of auxiliary scalar fields.

\section*{ACKNOWLEDGMENTS}

The work is supported by the National Key R$\&$D Program of China grants No. 2022YFF0503404 and No. 2022SKA0110100.

%\section*{References}
\newcommand\arctanh[3]{~arctanh.{\bf ~#1}, #2~ (#3)}
\newcommand\ARNPS[3]{~Ann. Rev. Nucl. Part. Sci.{\bf ~#1}, #2~ (#3)}
\newcommand\AL[3]{~Astron. Lett.{\bf ~#1}, #2~ (#3)}
\newcommand\AP[3]{~Astropart. Phys.{\bf ~#1}, #2~ (#3)}
\newcommand\AJ[3]{~Astron. J.{\bf ~#1}, #2~(#3)}
\newcommand\GC[3]{~Grav. Cosmol.{\bf ~#1}, #2~(#3)}
\newcommand\APJ[3]{~Astrophys. J.{\bf ~#1}, #2~ (#3)}
\newcommand\APJL[3]{~Astrophys. J. Lett. {\bf ~#1}, L#2~(#3)}
\newcommand\APJS[3]{~Astrophys. J. Suppl. Ser.{\bf ~#1}, #2~(#3)}
\newcommand\JHEP[3]{~JHEP.{\bf ~#1}, #2~(#3)}
\newcommand\JMP[3]{~J. Math. Phys. {\bf ~#1}, #2~(#3)}
\newcommand\JCAP[3]{~JCAP {\bf ~#1}, #2~ (#3)}
\newcommand\LRR[3]{~Living Rev. Relativity. {\bf ~#1}, #2~ (#3)}
\newcommand\MNRAS[3]{~Mon. Not. R. Astron. Soc.{\bf ~#1}, #2~(#3)}
\newcommand\MNRASL[3]{~Mon. Not. R. Astron. Soc.{\bf ~#1}, L#2~(#3)}
\newcommand\NPB[3]{~Nucl. Phys. B{\bf ~#1}, #2~(#3)}
\newcommand\CMP[3]{~Comm. Math. Phys.{\bf ~#1}, #2~(#3)}
\newcommand\CQG[3]{~Class. Quantum Grav.{\bf ~#1}, #2~(#3)}
\newcommand\PLB[3]{~Phys. Lett. B{\bf ~#1}, #2~(#3)}
\newcommand\PRL[3]{~Phys. Rev. Lett.{\bf ~#1}, #2~(#3)}
\newcommand\PR[3]{~Phys. Rep.{\bf ~#1}, #2~(#3)}
\newcommand\PRd[3]{~Phys. Rev.{\bf ~#1}, #2~(#3)}
\newcommand\PRD[3]{~Phys. Rev. D{\bf ~#1}, #2~(#3)}
\newcommand\RMP[3]{~Rev. Mod. Phys.{\bf ~#1}, #2~(#3)}
\newcommand\SJNP[3]{~Sov. J. Nucl. Phys.{\bf ~#1}, #2~(#3)}
\newcommand\ZPC[3]{~Z. Phys. C{\bf ~#1}, #2~(#3)}
\newcommand\IJGMP[3]{~Int. J. Geom. Meth. Mod. Phys.{\bf ~#1}, #2~(#3)}
\newcommand\IJTP[3]{~Int. J. Theo. Phys.{\bf ~#1}, #2~(#3)}
\newcommand\IJMPD[3]{~Int. J. Mod. Phys. D{\bf ~#1}, #2~(#3)}
\newcommand\IJMPA[3]{~Int. J. Mod. Phys. A{\bf ~#1}, #2~(#3)}
\newcommand\GRG[3]{~Gen. Rel. Grav.{\bf ~#1}, #2~(#3)}
\newcommand\EPJC[3]{~Eur. Phys. J. C{\bf ~#1}, #2~(#3)}
\newcommand\PRSLA[3]{~Proc. Roy. Soc. Lond. A {\bf ~#1}, #2~(#3)}
\newcommand\AHEP[3]{~Adv. High Energy Phys.{\bf ~#1}, #2~(#3)}
\newcommand\Pramana[3]{~Pramana.{\bf ~#1}, #2~(#3)}
\newcommand\PTEP[3]{~PTEP.{\bf ~#1}, #2~(#3)}
\newcommand\PTP[3]{~Prog. Theor. Phys{\bf ~#1}, #2~(#3)}
\newcommand\APPS[3]{~Acta Phys. Polon. Supp.{\bf ~#1}, #2~(#3)}
\newcommand\ANP[3]{~Annals Phys.{\bf ~#1}, #2~(#3)}
\newcommand\RPP[3]{~Rept. Prog. Phys. {\bf ~#1}, #2~(#3)}
\newcommand\ZP[3]{~Z. Phys. {\bf ~#1}, #2~(#3)}
\newcommand\NCBS[3]{~Nuovo Cimento B Serie {\bf ~#1}, #2~(#3)}
\newcommand\AAP[3]{~Astron. Astrophys.{\bf ~#1}, #2~(#3)}
\newcommand\MPLA[3]{~Mod. Phys. Lett. A.{\bf ~#1}, #2~(#3)}
\newcommand\NT[3]{~Nature.{\bf ~#1}, #2~(#3)}
\newcommand\PT[3]{~Phys. Today. {\bf ~#1}, #2~ (#3)}
\newcommand\APPB[3]{~Acta Phys. Polon. B{\bf ~#1}, #2~(#3)}
\newcommand\NP[3]{~Nucl. Phys. {\bf ~#1}, #2~ (#3)}
\newcommand\JETP[3]{~JETP Lett. {\bf ~#1}, #2~(#3)}
\newcommand\PDU[3]{~Phys. Dark. Univ. {\bf ~#1}, #2~(#3)}


\begin{thebibliography}{99}
 \bibitem{picon:1999}C. Armendariz-Picon, T. Damour, and V. Mukhanov,
\PLB{458}{209}{1999}.
\bibitem{garriga:1999} J. Garriga and V. F. Mukhanov, \PLB{458}{219}{1999}.
[arXiv:hep-th/9904176].
\bibitem{chiba:2000} T. Chiba, T. Okabe, and M. Yamaguchi, \PRD{62}{023511}{2000}.
\bibitem{picon:2000} C. Armendariz-Picon,V. Mukhanov, and P. J. Steinhardt,
\PRL{85}{4438}{2000}.
\bibitem{picon:2001} C. Armendariz-Picon,V. Mukhanov, and P. J. Steinhardt, \PRD{63}{103510}{2001}.
\bibitem{chiba:2002} T. Chiba, \PRD{66}{063514}{2002}..
\bibitem{chimento:2004} L. P. Chimento and A. Feinstein, \MPLA{19}{761}{2004}.
\bibitem{chimento:2003} L. P. Chimento, \PRD{69}{123517}{2004}, astro-ph/0311613
\bibitem{malquarti:2003} M. Malquarti, E. J. Copeland and Andrew R. Liddle,\PRD{68}{023512}{2003}.


\bibitem{robert:2004} R. J. Scherrer, \PRL{93}{2004}{011301}.   
\bibitem{chimento:2005} L. P. Chimento and R. Lazkoz, \PRD{71}{023505}{2005}.
\bibitem{abramo:2006} L. R. Abramo, N. Pinto-Neto, \PRD{73}{063522}{2006}.
\bibitem{bonvin:2006} C. Bonvin, C. Caprini and R. Durrer, \PRL{97}{081303}{2006}.
\bibitem{bamba:2012} K. Bamba, J. Matsumoto and S. Nojiri, \PRD{85}{084026}{2012}.


\bibitem{jean:2007} J. P. Bruneton, \PRD{75}{085013}{2007}. 

\bibitem{chiba:2009}  T. Chiba, S, Dutta, and R. J. Scherrer, \PRD{80}{043517}{2009}.


\bibitem{nojiri:2019}
S. Nojiri, S. D. Odintsov and V. K. Oikonomou,
\NPB{941}{11-27}{2019}.

\bibitem{odintsov:2020}
S. D. Odintsov and V. K. Oikonomou,
\CQG{37}{025003}{2020}.

\bibitem{odintsov:2020pdu}
S. D. Odintsov, V. K. Oikonomou and F. P. Fronimos,\PDU{29}{100563}{2020}.


\bibitem{herrera:2020}  R. Herrera, \PRD{102}{123508}{2020}.
\bibitem{gia:2020} I. D. Gialamasa and A. B. Lahanasa, \PRD{101}{084007}{2020}.

\bibitem{odintsov:2021}
S. D. Odintsov, V. K. Oikonomou and F. P. Fronimos,
\NPB{963}{115299}{2021}.

 \bibitem{bose:2009a} N. Bose and A. S. Majumdar, \PRD{79}{103517}{2009}.
\bibitem{bose:2009b}  N. Bose and A. S. Majumdar, \PRD{80}{103508}{2009}.


\bibitem{tian:2021}  S. X. Tian and Z.-H. Zhu, \PRD{103}{043518}{2021}.

\bibitem{riess:2021} A. G. Riess, S. Casertano, W. Yuan, J. Bradley Bowers, L.
Macri, J. C. Zinn, and D. Scolnic, \APJL{908}{6}{2021}.
\bibitem{freedman:2021}  W. L. Freedman, \APJ{919}{16}{2021}.

\bibitem{lee:2022} Bum-Hoon Lee et al, \JCAP{04}{004}{2022}. 


\bibitem{hung:2023} Y. S. Hung and S. P. Miao, \PRD{107}{103533}{2023}.

\bibitem{kehayias:2019}J. Kehayias, R. J. Scherrer, \PRD{100}{023525}{2019}.


\bibitem{sarkar:2023} A. Sarkar, A. Ali, K. R. Nayak, A. S. Majumdar, \PRD{107}{084038}{2023}.

\bibitem{lara:2022} G. Lara, M. Bezares, E. Barausse, \PRD{105}{064058}{2022}.

\bibitem{huang:2021}Z. Huang, \PRD{104}{103533}{2021}.

\bibitem{chatt:2021} A. Chatterjee, S. Hussain, K. Bhattacharya, \PRD{104}{103505}{2021}


\bibitem{nagy:2021} C. Nagy, Z. Keresztes, L. Á. Gergely, \PRD{103}{124056}{2021}.

\bibitem{barvinsky:2021} A. O. Barvinsky, N. Kolganov, and A. Vikman,  \PRD{103}{064035}{2021}. 

\bibitem{fer:2024} A. L. Ferreira, J. N. Pinto-Neto and J. Zanelli, \PRD{109}{023515}{2024}.
\bibitem{langlois:2008} D. Langlois and S. Renaux-Petel, \JCAP{0804}{017}{2008}. 

\bibitem{gao:2022}C. Gao, arXiv: gr-qc/2203.05957.

\bibitem{pad:2003}T. Padmanabhan, \PR{380}{235}{2003}.

\bibitem{afshordi:2007}N. Afshordi, D. J. H. Chung, and G. Geshnizjani, \PRD{75}{083513}{2007}, arXiv:hep-th/0609150.
\end{thebibliography}
\end{document}